\documentclass[aip,a4paper,reprint,groupedaddress]{revtex4-1}

\usepackage{graphicx}
\usepackage{siunitx}
\usepackage{amsmath}
\usepackage{hyperref}
\usepackage{enumitem}
\usepackage{color}

\hypersetup{
	colorlinks   = true,
	urlcolor     = blue,
	linkcolor    = blue,
	citecolor   = blue
}

\DeclareSIUnit\Molar{\textsc{m}}

\newcommand{\ms}[1]{\SI{#1}{\milli \second}}

\begin{document}
\title{Bayesian inference for nanopore data analysis}

\author{Niklas Ermann}
\author{Kaikai Chen}
\author{Ulrich F. Keyser}
\email{ufk20@cam.ac.uk}
\affiliation{Cavendish Laboratory, University of Cambridge, 19 JJ Thomson Avenue, Cambridge CB3 0HE, UK}

\begin{abstract}
	Nanopore sensors detect the substructure of individual molecules from modulations in an ion current as molecules pass through them. In this work, we present the classification of features in the substructure as a case study to illustrate the power of Bayesian inference when analysing nanopore data. A brief introductory section provides an overview of the core concepts, followed by a detailed description of the analysis procedure to facilitate other researchers to add Bayesian inference to their toolbox. Our hybrid approach of a classical peak-finding algorithm and Bayesian model comparison allows the probabilistic classification of features as ``0'' or ``1'' bits by calculating relative evidences for two competing models. We correctly classify on average $\sim 70\%$ of bits for individual events and use the probabilistic nature of the approach to calculate a cumulative estimate with an accuracy of $> 94\%$. The technique presented here is readily extensible to models of the translocation process which can take into account arbitrary molecular designs, our approach may therefore be used to analyse a wide range of features observed in nanopore experiments.
\end{abstract}


\maketitle

\section{Introduction}

\begin{figure}
	\includegraphics{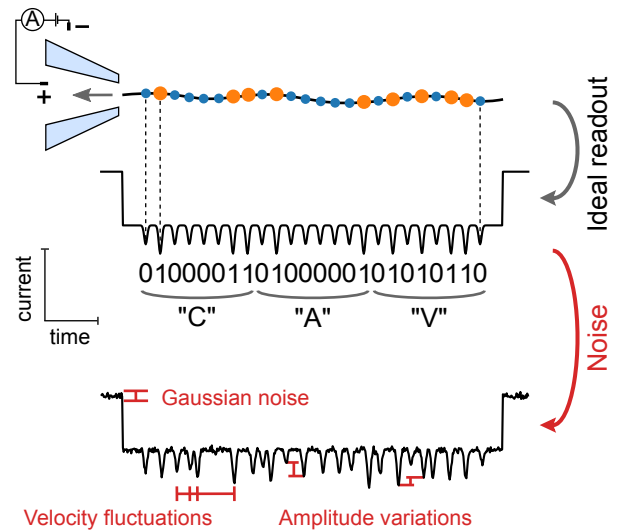}
	\caption{Experimental noise complicates the readout of bit sequences stored on a double-stranded DNA scaffold. In the ideal case, bits can be assigned directly from the amplitude of the secondary current drops produced by modifications. The upper idealised current trace shows that the letters ``CAV'' can be easily decoded in 8-bit ASCII. The simulated bottom trace demonstrates that Gaussian noise inherent in the measurement, velocity fluctuations of the DNA strand and variations in peak amplitudes complicate the decoding.}
	\label{fig:nanopore}
\end{figure}

Resistive-pulse sensing based on nanopores is utilised not only in the detection but increasingly in the study of substructure of single molecules \cite{Waduge2017, Loh2018}. Characterisation of molecular substructure relies on accurately identifying features \textit{within} a translocation signal. For example, such features in the ionic current could correspond to DNA folds \cite{Li2003}, knots \cite{Plesa2016} or modifications such as protrusions \cite{Plesa2015, Bell2016} or attached proteins \cite{Raillon2012a}. Modifications offer the possibility to store digital data \cite{Chen2018} or indicate the presence of certain base sequences \cite{Smeets2009}. In the example shown in Figure \ref{fig:nanopore}, each modification produces a secondary current drop in the already reduced current, creating a bit sequence of high ``1'' and low ``0'' signals which can be used to encode information. In the ideal scenario, the amplitude of a drop would relate exactly to the bit encoded at this specific position. In a real nanopore system, however, several sources of noise complicate the decoding process \cite{Smeets2008}. The bottom part of Figure \ref{fig:nanopore} illustrates that fluctuations in the velocity of a translocating DNA strand \cite{Storm2005a} shift the position of secondary drops, preventing the exact localisation of bits purely based on their appearance in the signal. Variations in the drop amplitude create ambiguity as to which modification produced the feature, which is exacerbated by the Gaussian noise inherent in the current measurement \cite{Smeets2008}. As a result, analysis procedures which aim to extract the molecular substructure that gives rise to the features within a translocation signal need to be able to distinguish between signal and noise and quantify the probability that a certain structure has been detected.

Existing approaches to nanopore data analysis have been dominated by classical algorithms. Several techniques have focused on finding the true magnitude and duration of the current drop when constrained by the finite filter rise times inherent in instrumentation \cite{Pedone2009, Balijepalli2015}. Raillon et al. developed a cumulative sums algorithm to identify sequential steps during a translocation \cite{Raillon2012}. Further efforts went into the detection of secondary current drops as well as the development of comprehensive software tools for the analysis of nanopore data \cite{Bell2016, Plesa2015a, Forstater2016}. While these classical techniques can provide fast and reliable event characterisation for their specific use case, their performance often relies on user-defined inputs and arbitrary thresholds. Recent work on convolutional neural networks has shown that they successfully classify events in a generalised fashion, at superior data usage and accuracy compared to conventional techniques \cite{Misiunas2018}. However, the approach requires labelled datasets containing thousands of events for each molecular species to be detected.

Bayesian inference is a promising alternative for the extraction of information from experimental data \cite{Gelman1995}. In the past, its use has been hampered by the computational cost associated with calculating high-dimensional integrals. This hurdle has been partially overcome by advances in computational power over the last decades as well as the development of efficient Markov chain Monte Carlo-based (MCMC) algorithms \cite{Hines2015}. The probabilistic nature of Bayesian techniques allows the calculation of probability distributions for model parameters, as opposed to the single-value estimates often found in classical approaches. The probability landscapes directly show a wealth of features which may otherwise remain obscured, such as multimodal distributions and regions surrounding maximum-likelihood parameter estimates \cite{Sivia2006}. In addition to the characterisation of model parameters, the Bayesian approach assigns a probability to the validity of the model itself. Called evidence, this value allows comparing different candidate models to assess which one best explains the experimental data and thus which physical theory should be favoured \cite{MacKay}.

In this paper, we demonstrate that the probabilistic nature of Bayesian inference allows us to draw unbiased conclusions about the substructure of translocating molecules from noisy current data. As a case study, we use recently published data from a double-strand of DNA which has been modified with differently sized DNA overhangs at defined positions along its length \cite{Chen2018}. As shown in Figure \ref{fig:nanopore}, each modification produces a secondary current drop in addition to the already reduced current whose magnitude depends on the length of the overhang. The total number of overhangs is defined through the design of the molecule, however the bit sequence remains unknown. The question for data readout thus becomes: given a number of events, what is the relative probability of an overhang at a certain position being long or short, i.e. signifying a ``0'' or ``1'' bit? In the following we show that a Bayesian approach is ideally suited to analyse substructure in nanopore translocation data. Before illustrating the practical procedure, we give a brief introduction into Bayesian inference and reasoning.

\section{Bayesian inference}
\label{sec:bayes}

Bayesian techniques emerge from a fundamental statement about conditional probabilities called Bayes' rule \cite{Canton1763}:

\begin{align}
\label{eq:bayesrule}
	\underbrace{\strut p(\theta|D,M)}_{\textrm{posterior}} = \frac{\overbrace{\strut p(D|\theta, M)}^{\textrm{likelihood}} \overbrace{\strut p(\theta|M)}^{\textrm{prior}}}{\underbrace{\strut p(D|M)}_{\textrm{evidence}}}
\end{align}

Here $\theta$ denotes the model parameters, $D$ the experimental data and $M$ a certain model. Bayes' rule tells us that the probability for the model parameters to take on a certain value given the data, called the posterior, is proportional to the probability of obtaining the data from the given choice of parameters, called the likelihood. While the two probabilities sound similar, this is in fact a powerful statement: in data analysis we are interested in the posterior, i.e. we want to know what the data can tell us about the parameters of our explanatory model. Bayes' rule allows us to obtain the posterior by relating it to a probability we can calculate, the likelihood. Using a nanopore signal as an example, Figure \ref{fig:nanopore} shows that the current trace exhibits Gaussian noise whose variance we call $\sigma^2_n$. We represent any model $M$ describing the data with a function $h(\theta)$ which defines how the model relates its parameters $\theta$ to the observed signal. The likelihood for an individual data point $d$ given $\theta$ and $M$ is then simply proportional to the Gaussian noise probability distribution centered around $d$ at the point predicted by $h(\theta)$, that is

\begin{align}
\label{eq:gaussnoise}
	p(d|\theta, M) \propto \exp\bigg(-\frac{(h(\theta)-d)^2}{2 \cdot \sigma_n^2}\bigg)
\end{align}

Having obtained an expression for the likelihood, the remaining term with a dependence on the parameters $\theta$ is the prior $p(\theta|M)$. It encodes previous knowledge about the parameters, for example from the posteriors obtained from previous experimental data \cite{Sivia2006}. A lack of initial information about $\theta$ is easily represented by a flat probability distribution. The prior in combination with the likelihood lets us calculate the quantity of interest, namely the posterior probability over the parameter values given the data. An accessible treatment of this approach can be found in \cite{Sivia2006}, including examples on how the Bayesian view motivates widespread statistical techniques such as least squares regression.

Up to this point, the discussion has focused on determining the parameter probability distribution for a given model. However, within the probabilistic framework we can equally ask which of several models is most likely given a set of experimental data. To answer this question we need to calculate $p(M_i|D)$ where $i$ indexes several competing models. Bayes' rule tells us that

\begin{align*}
	p(M|D) = \frac{p(D|M) p(M)}{p(D)} \propto p(D|M)
\end{align*}

In the last term of the equation we recognise the evidence $p(D|M)$ which appears in Equation \ref{eq:bayesrule}. Assuming uniform priors over different models $M_i$, the model with the highest probability of explaining the data is thus the one with the largest evidence. To compute the evidence, we note that

\begin{align}
\label{eq:evidence}
	p(D|M) &= \int p(D, \theta|M) d\theta \\
	&= \int \underbrace{\strut p(D|\theta,M)}_\textrm{likelihood} \underbrace{\strut p(\theta|M)}_\textrm{prior} d\theta
\end{align}

The evidence is therefore an integral of likelihood and prior over the entire parameter space. Particularly for high-dimensional models where analytical optimisations aren't possible, it is these integrals which make Bayesian techniques computationally intensive. However, once the calculation becomes feasible we obtain a probabilistic method to assess the explanatory power of different models. In the next section we show how this Bayesian tool allows us to extract unbiased information from nanopore data. The interested reader is referred to \cite{Gelman1995} and \cite{MacKay} for more detailed discussions of Bayesian inference.

\section{Case study: bit sequence identification from nanopore data}

\begin{figure}
	\includegraphics{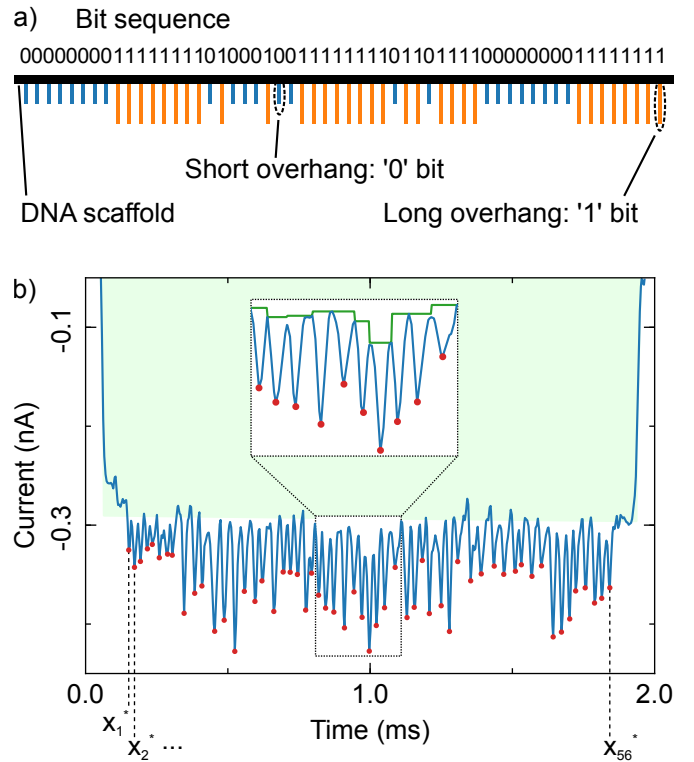}
	\caption{Modifications along a double-stranded DNA scaffold encode 56 bits, as published in \cite{Chen2018}. \textbf{(a)} Schematic of the DNA molecule. Short DNA overhangs along the scaffold's length encode ``0'' bits (blue), long overhangs represent ``1'' (orange). \textbf{(b)} An example current trace produced by the molecule in (a). Red dots indicate the secondary current drops selected by a classical peak-finding algorithm, each drop corresponds to one of the 56 overhangs. We determine the drop amplitudes relative to the shaded green region, which is calculated from an auxiliary trace where the drops have been removed (continuous green line in the inset).}
	\label{fig:one_event}
\end{figure}

As mentioned in the introduction, we use previously published data from a modified double-strand of DNA to illustrate how nanopore data analysis can benefit from Bayesian techniques \cite{Chen2018}. Each of the overhangs along the strand's length produces a secondary current drop whose amplitude depends on the overhang's size (Figure \ref{fig:one_event}). In total the strand has been decorated with 56 such modifications, resulting in a sequence of 56 bits. For each bit, the goal is to estimate whether it corresponds to a ``0'' or ``1'', depending on the size of the overhang. The experimental data was obtained from a homogeneous sample with each strand modified in the same way. This means that we can sequentially build up the estimate for each bit as more events are added to the total dataset. We denote the data for a single event $i$ by $D_i$ and the total dataset encompassing all events by $\{D_i\}$. Each $D_i$ consists of current data $y$ and time points $x$.

We initially carry out several preprocessing steps. These use classical algorithms and return the intra-event times and amplitudes of the 56 most significant secondary current drops. We discuss the procedure in the following to present the complete analysis pipeline.

\subsection{Preprocessing steps}
\label{subsec:preproc}

The goal of the initial data processing is to provide values for the priors used in the comparison of the ``0'' bit and ``1'' bit models in the next section. First, we determine current levels corresponding to the unimpeded pore $L_0$ and one double strand in the pore $L_1$. This is achieved by compiling the concatenated current data from all events into a kernel density estimate and detecting the two peaks in the distribution as described in \cite{Ermann2018a}. We define a cut-off value $C$ as $C = L_0 - 0.75*(L_0-L_1)$ which will be used later to define the start and end of each event.

Next, we carry out the following steps for each event $D_i$ with current trace $y$ and time points $x$:

\begin{enumerate}
	\item Secondary current drop identification
	
		We detect the secondary current drops for each event using a peak-finding algorithm adapted from \cite{Plesa2015a}. We are tolerant of false positives at this point and may obtain more than 56 drops. These will be filtered out in a later step. We discard events for which the peak finding results in fewer than 56 drops.
	
	\item \label{peakfinding} Identification of intra-event baseline current.
	
	We use programmatic notation in the following, where $\bar{v}[i:j]$ refers to the slice of vector $\bar{v}$ between indices $i$ and $j$. Assignments of the form $A = B$ work from right to left, i.e. assign the value $B$ to the variable $A$.
	
	\begin{enumerate}
		\item We first remove the secondary current drops identified in the previous step by setting the current values in each subregion found to be part of a current drop to the most positive value within that subregion. This gives us a new current trace $\hat{y}$ with the drop section replaced by piecewise constant values as shown by the green line in the inset in Figure \ref{fig:one_event} b).
		\item On the modified current trace $\hat{y}$ we determine two index values $a$ and $b$ which define the first and last positions at which $\hat{y} < C$, where $C$ is the cut-off value calculated initially. We then fit a linear function $f(x,s,i) = i + s*x$ to the region $\hat{y}[a:b]$, obtaining least-squares fit values $\tilde{i}$ and $\tilde{s}$.
		\item On the original current trace $y$, we again determine two index values $a$ and $b$ corresponding to the first and last positions at which $y < C$. We then replace the slice between these indices with values calculated from the function $f(x,\tilde{s}, \tilde{i})$ from the previous step, i.e. if the new current trace $\tilde{y}$ is initially the same as $y$, we assign $\tilde{y}[a:b] = f(x[a:b],\tilde{i},\tilde{s})$ to obtain a new current trace $\tilde{y}$ that contains a linear intra-event baseline between the start and end of the event without taking into account secondary drops. This trace, shown by the shaded green region in Figure \ref{fig:one_event} b), serves as the baseline from which we calculate the amplitudes of the secondary drops.
	\end{enumerate}
	
	\item Drop filtering
	
		In the case of more than 56 drops, we keep the 56 with the largest distance from the baseline $\tilde{y}$. In the example event in Figure \ref{fig:one_event} b) the selected drops are marked with red dots. This simple filtering step makes the system susceptible to shift errors, which occur when spurious drops shift the position assignments of all other bits. Such errors as well as techniques to avoid them are discussed in section \ref{subsec:bayes}.
		
\end{enumerate}
		
Having identified drops for each event, we calculate aggregate values across the whole dataset $\{D_i\}$ excluding events for which fewer than 56 drops could be detected. We distinguish two populations in the drop amplitude corresponding to ``0'' and ``1'' bits. By fitting Gaussians to the peaks in the kernel density estimate of all drop amplitudes within an event we obtain estimates of the mean secondary drop amplitude $m_j$ and the spread in amplitudes $w_j$ where $j=0,1$ for ``0'' and ``1'' bits respectively. It should be noted that the two populations can only be clearly distinguished for some events, we use the arithmetic mean for all events for which identification is possible to obtain the average mean and width of the amplitude drops for each nanopore, $\bar{m}_j$ and $\bar{w}_j$.

\subsection{Bayesian model comparison}
\label{subsec:bayes}

The above preprocessing steps provide us with the positions of the most prominent 56 current drops for each event, as well as the distribution of drop amplitudes across the whole dataset. We use a classical algorithm to identify current drops, as opposed to running Bayesian inference on a large model that fully incorporates all features of a translocation event. The classification of 56 bits means such a model would contain at least 56 parameters. State-of-the-art nested sampling algorithms that allow full Bayesian inference scale at worst as $E \sim \mathcal{O}(P^3)$ or even exponentially, where $E$ is the number of likelihood evaluations and $P$ is the number of model parameters \cite{Handley2015}. A large number of parameters thus strongly increases the computation time, which makes the full model approach unfeasible for a higher number of bits. The computational cost is exacerbated by the variation in the times between consecutive drops due to fluctuations in the translocation velocity of the DNA strand. Designing ``1'' and ``0'' bits as differently sized overhangs as opposed to the presence and absence of modifications somewhat mitigates this issue and leads to a greatly improved ability to identify current drops. However, attempts to create a full model showed that a large number of additional parameters is still required to correctly detect drops when the distance between them cannot be constrained to one value.

\begin{figure*}
	\includegraphics{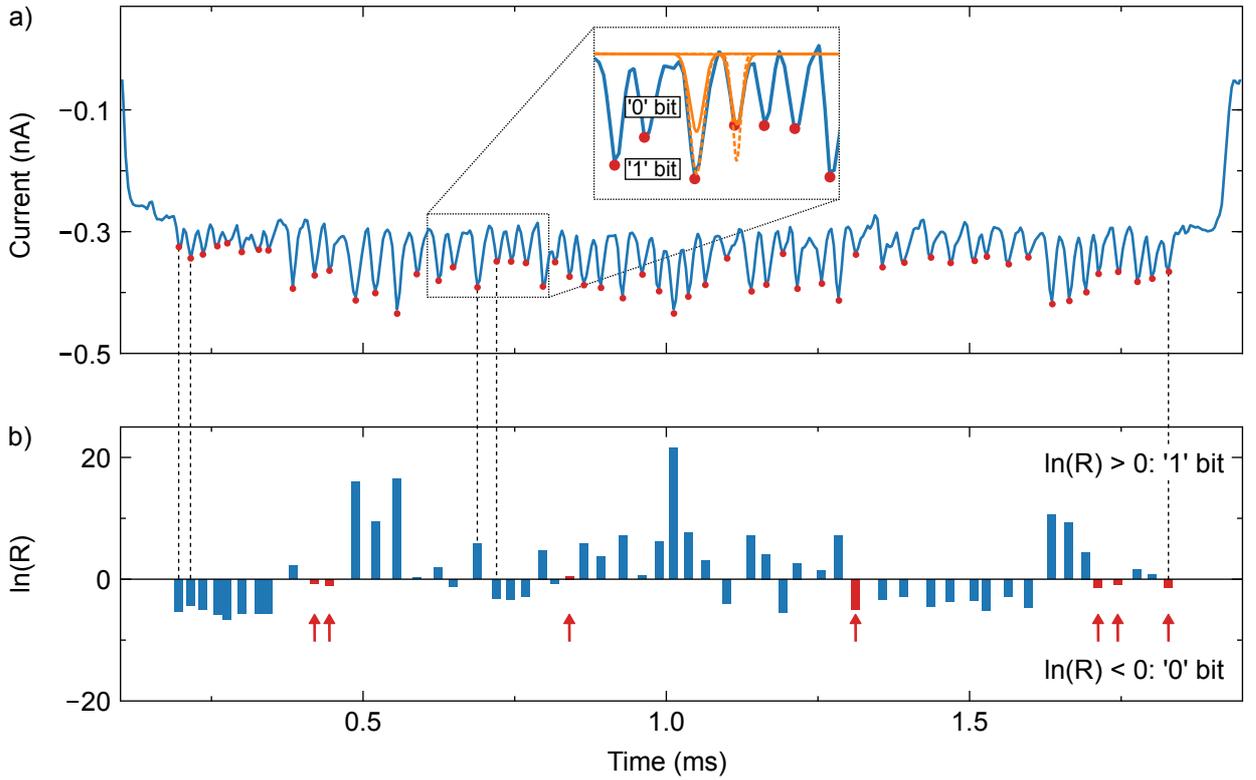}
	\caption{Bayesian model comparison assigns bits to each of the 56 current drops in the single event shown. \textbf{(a)} Illustration of the analysis procedure on the translocation event from Figure \ref{fig:one_event} (b). Stepping through the identified current drops, we compare the evidences for two models representing a ``0'' and ``1'' bit. The orange traces show the model outcomes for maximum likelihood parameter estimates on two exemplary drops, continuous and dashed lines correspond to the ``0'' and ``1'' bit models respectively. It is clear that the ``1'' model more accurately describes the data for the first drop, whereas the ``0'' model fits the second drop. \textbf{(b)} Outcome of the Bayesian model comparison. $\ln(R)$ describes the logarithm of the ratio of the evidences for the ``1'' and ``0'' models, $\ln(R) > 0$ indicates a ``1'' bit and vice versa. The analysis correctly classifies 49 out of 56 bits in the event shown, red bars and arrows show wrong assignments.}
	\label{fig:analysis}
\end{figure*}

For the above reasons we follow a hybrid approach by using a classical algorithm to identify the 56 most prominent current drops and Bayesian inference to consecutively assign a bit to each drop, which allows us to take advantage of Bayesian probabilities without excessive computational demands. To do so, we use the information obtained in the preprocessing steps to define two models, $M_0$ and $M_1$, corresponding to a single ``0'' and ``1'' bit respectively. Within each event, we then use the Bayesian approach to compare the models' evidence values, thereby classifying each current drop as ``0'' or ``1''.

From Equation \ref{eq:evidence} it is apparent that we require an expression for the likelihood to calculate evidences. Assuming Gaussian noise in the current signal, we define the likelihood as

\begin{align*}
	P(D|\theta,M_j) \propto \exp\bigg(-&\sum_{k=1}^L\frac{(h(x_k,\tilde{y}_k,\theta)-y_k)^2}{2 \cdot \sigma_n^2}\bigg)
\end{align*}

where $x_k$, $y_k$ and $\tilde{y}_k$ are the $k$th elements of the time and the initial and corrected current data respectively and the second line is directly comparable to Equation \ref{eq:gaussnoise}. The sum over the trace elements stems from the assumption of independent samples, meaning we can multiply the Gaussian noise distributions for each element.

Within the exponent, $L$ is the length of the current trace and $\sigma_n$ the Gaussian noise in the current. The function $h(x,b,\theta)$ fully defines how the model $M_j$ relates its parameters $\theta$ to measured data. We represent the secondary current drops by Gaussian functions with mean $\mu$, amplitude $a$ and width $\sigma$ added to a baseline $b$, i.e.

\begin{align*}
	h(x,b,\theta) &= h(x,b,\mu,a,\sigma) \\
	&= a \cdot \exp\bigg(-\frac{(x-\mu)^2}{2 \cdot \sigma^2}\bigg) + b
\end{align*}

In the analysis presented here, this model is used for both ``0'' and ``1'' bits, with the distinction between the two coming from different priors on the amplitude parameter $a$. This is an empirical simplification as a detailed physical understanding of how modifications produce their associated current drops is the subject of ongoing research. In addition, at this point it is not known to what extent the differently sized overhangs influence other drop characteristics such as shape. As more is understood about the underlying process and the experimental data the two models should be refined.

In addition to the likelihood, we require prior distributions over the model parameters $\theta$. We choose uniform, normalised priors, as described in the appendix to \cite{Handley2015}. We constrain the values for the mean $\mu$ for the $l$th current drop to the interval $[x_l^*-0.5 \cdot \overline{\Delta x^*}, x_l^*+0.5 \cdot \overline{\Delta x^*}]$, where $x_l^*$ is the time position of the $l$th drop and $\overline{\Delta x^*}$ is the mean difference between all drop positions within the event (see Figure \ref{fig:one_event}). A prior in the interval $[\ms{0.002}, \ms{0.006}]$ accurately describes the width $\sigma$ of the secondary current drops. As mentioned above, the distinction between the two hypotheses comes from different priors on the amplitude $a$ of the Gaussian function:

\begin{alignat*}{3}
	&P(a|M_j) \\
	&=\begin{cases}
	(a_{j,\textrm{max}} - a_{j,\textrm{min}})^{-1}&{\text{for }}a_{j,\textrm{min}} < a < a_{j,\textrm{max}} \\
	0&{\text{otherwise.}} \\
	\end{cases} \\[12pt]
	&\textrm{where } \\
	&a_{j,\textrm{min}} = \bar{m}_j + s_{j,\textrm{min}} \cdot \bar{w}_j \\
	&a_{j,\textrm{max}} = \bar{m}_j + s_{j,\textrm{max}} \cdot \bar{w}_j \\
\end{alignat*}

and $\bar{m}_j$ and $\bar{w}_j$ are the mean and width of the current drop populations identified at the end of section \ref{subsec:preproc}. The factors in the prior limits were chosen as $s_{0,\textrm{min}} = -0.5$, $s_{0,\textrm{max}} = +0.5$, $s_{1,\textrm{min}} = -2$ and $s_{1,\textrm{max}} = 0$ to optimally represent the current drop amplitudes for the two bits $j = 0,1$. A simple Gaussian fit to each current drop and classification according to an amplitude threshold would be less computationally intensive. However, as mentioned above the approach presented here is generic in that it can easily be extended to take into account current drop characteristics such as shape by adapting the model function $h(x,b,\theta)$. Secondly, we obtain evidences for each model, which means the classification into ``0'' and ``1'' bits is easily combined into an aggregate value across many events.

Having obtained the likelihood and prior, we use the \textit{MultiNest} Bayesian inference algorithm to compute log evidence values $\ln \big(P(D|M_j)\big)$ \cite{Feroz2009}. The inset in Figure \ref{fig:analysis} a) illustrates our sequential approach to bit assignment for two exemplary bits: at the position of each bit, we calculate evidences for the competing models, select the classification with the higher relative probability and move on the next position. The orange traces show the model outcomes using maximum likelihood parameter estimates which we obtain as part of the evidence calculation. It is clear that the ``1'' model matches the first current drop more accurately, whereas the ``0'' model is preferred for the second drop. The Bayesian approach represents the concept of a `better match' by the ratio of the posterior probabilities over the two models, $R$, which is related to the log evidences via

\begin{align*}
	\ln \big(R\big) &= \ln \bigg( \frac{P(M_1|D)}{P(M_0|D)} \bigg) = \ln \bigg( \frac{P(D|M_1)P(M_1)}{P(D|M_0)P(M_0)} \bigg) \\
	&= \ln \bigg( \frac{P(D|M_1)}{P(D|M_0)} \bigg) \\
	&= \ln \big(P(D|M_1)\big) - \ln \big(P(D|M_0)\big)
\end{align*}

where in the second line we assume equal priors for the two models, i.e. $P(M_1) = P(M_0)$. For each current drop, we therefore calculate the difference between the log evidences for a ``1'' bit and a ``0'' bit to obtain the log of the probability ratio. The resulting value $\ln(R)$ provides both a decision criterion between the two bits and an indication how strongly one hypothesis is favoured: $\ln(R)<(>)0$ indicates a preference for the ``0'' (``1'') bit, while the absolute value $|\ln(R)|$ represents the preference strength. Figure \ref{fig:analysis} b) shows the outcome of the calculation for all bits in a single event. Our approach misclassifies 7 out of 56 bits for this particular example as shown by the red bars and arrows.

Analysis of more events showed that shift errors are a common source of wrong bit assignments. These occur when the classification of bits is correct, but the erroneous insertion or deletion of bits lead to the wrong assignment of subsequent bits to positions in the sequence. Such errors are particularly problematic when the goal is to estimate the exact bit sequence for single events. If the goal of the DNA modification is to create a library of different sequences, a possible mitigation strategy is to not use all $56$ available dimensions but create sequences with a maximal distance in vector space. For an estimated bit sequence shift errors can then be taken into account to find the most likely classification. A more direct approach is to design overhangs so that they appear in bit ``blocks'' with detectable spacing between them, which limits the effect of erroneously identified current drops to the length of the block. Similarly, larger ``beacon'' overhangs allow verification of the position at certain intervals.

An additional way to improve the classification accuracy is to combine the bit estimates from multiple events. One of the advantages of the Bayesian approach is that it directly allows updating the bit estimate for each of the 56 positions as new events are added. As the DNA molecule can enter with either of its two ends first, we orient each event based on the number of ``0'' bits found in the 8 first and 8 last positions in the estimated bit sequence. It should be noted that apart from this correction, our classification includes no prior information on the bit sequence.

If $\{D_i\}$ is again the dataset of all events and $N$ is the number of events, the overall evidence for hypothesis $j$ is given by

\begin{align*}
	P\Big(\{D_i\}|M_j\Big) = \prod_{i=1}^{N}P\Big(D_i|M_j\Big)
\end{align*}

\begin{figure}
	\includegraphics{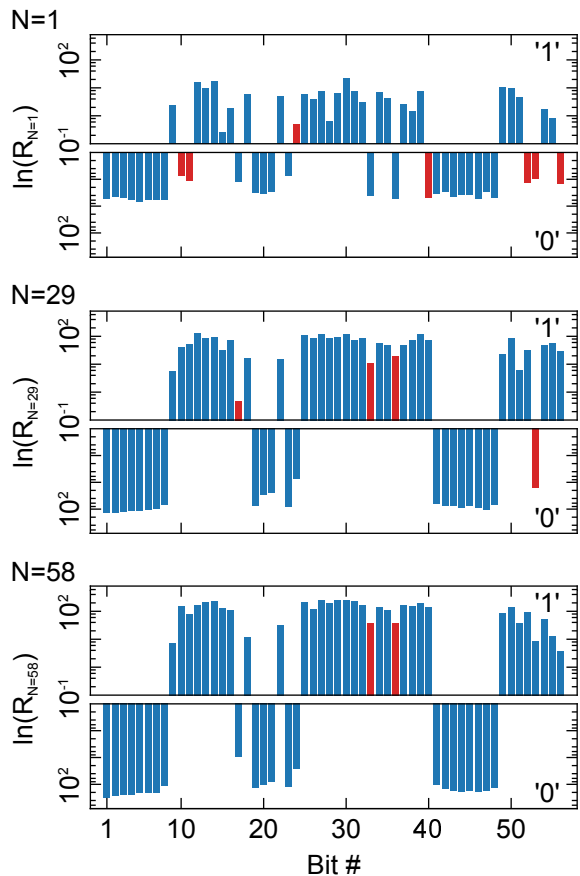}
	\caption{Combining bit estimates from multiple events improves the accuracy and confidence of assignments. $\ln (R_{N})$ describes the logarithm of the evidence ratio for the two models ``0'' and ``1'' compiled from $N$ events, here $N=1$, $N=29$ and $N=58$ from top to bottom. $\ln (R_{N}) > 0$ again indicates a ``1'' bit and vice versa, the magnitude $\ln (R_{N})$ shows how strongly one model is favoured. The number of wrong assignments (red bars) drops from 7 to 2 and the confidence increases as more event are included.}
	\label{fig:agg_confidence}
\end{figure}

A derivation of this result can be found in section I of the supplementary information. The cumulative probability ratio $R_{N}$ taking into account $N$ events is then given by

\begin{align*}
	&\ln \big(R_{N}\big) = \ln \bigg( \frac{P(M_{1}|\{D\})}{P(M_{0}|\{D\})} \bigg) \\
	&= \ln \Bigg( \frac{\prod_{i=1}^{N}P(D_i|M_{1})}{\prod_{j=1}^{N}P(D_j|M_{0})} \Bigg) = \ln \Bigg(\prod_{i=1}^{N} \frac{P(D_i|M_{1})}{P(D_i|M_{0})} \Bigg) \\
	&= \sum_{i=1}^{N}\Big(\ln \big(P(D_i|M_{1})\big) - \ln \big(P(D_i|M_{0})\big) \Big)
\end{align*}

where we again assumed equal priors for the two hypotheses in the second line. This means that to obtain a cumulative log probability ratio for a number of events we simply add up their individual log probability ratios. The decision criterion remains the same as in the individual case, where $\ln(R_{N}) < 0$ indicates a preference for the ``0'' bit and vice versa, while the absolute value $|\ln(R_{N})|$ represents how much one bit is favoured over the other. Figure \ref{fig:agg_confidence} shows the aggregate log probability ratios for each bit position calculated from 1, 29 and 58 events. The inclusion of more experimental data decreases the error rate to 2 out of 56 bits, misclassifications are again shown by the red bars. The magnitude of the bars increases with $N$, indicating that the addition of further events leads to more confident estimates. It should be noted that wrong bit assignments depend on how many and which events have been included. The wrong or low-confidence estimates occur mainly in positions where a change occurs from ``0'' to ``1'' or vice versa. This can be explained by the shift errors outlined above: wrongly assigning current drops to shifted positions in the bit sequence only produces an error if the neighbouring bit has a different value. The probability ratios obtained from the Bayesian approach thus directly indicate which bits are more difficult to classify, thereby pointing to the most likely source of errors and informing the design of improved DNA structures.

\begin{figure}
	\includegraphics{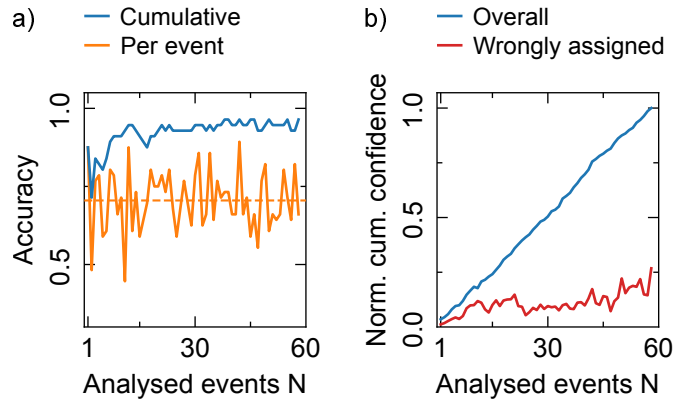}
	\caption{Analysing additional events improves the assignment accuracy as well as the confidence in bit estimates. \textbf{(a)} Assignment accuracy as a function of the number of analysed events $N$. While the individual (per event) accuracy fluctuates around $70\%$, the value for the cumulative estimate rises to above $90\%$ within a few events. \textbf{(b)} Cumulative confidence as measured by the mean absolute log probability ratio $\langle|\ln(R_{N})|\rangle$. The blue line shows that the confidence including all bit estimates increases steadily as we analyse more events. The confidence calculated by including only \textit{wrong} bit assignments, however, remains at a low level (red line).}
	\label{fig:convergence}
\end{figure}

To assess how the number of analysed events influences the error rate, Figure \ref{fig:convergence} a) shows the assignment accuracy for individual events as well as for the estimate based on the cumulative log probability ratio $\ln (R_{N})$. While the individual accuracy lies around $70\%$ on average (dashed orange line), the cumulative accuracy rises to above $90\%$ within a few events and remains unaffected by consecutive low-accuracy estimates (blue line). Figure \ref{fig:convergence} b) illustrates how the confidence in the assignments as measured by the mean absolute log probability ratio $\langle|\ln(R_{N})|\rangle$ increases steadily as we analyse more events (blue line). Crucially, however, the confidence in \textit{wrong} estimates does not rise substantially as more events are added (red line). This means that the increase in overall confidence is driven by more confident estimates mainly for correctly assigned bits, while estimates for wrongly classified bits remain uncertain. The Bayesian approach presented here thus correctly identifies large parts of the bit sequence while flagging wrong estimates with a low confidence.

\section{Conclusion}

We have demonstrated that Bayesian inference is a powerful technique to analyse current data obtained from nanopore measurements. Using the readout of a bit sequence encoded as modifications on a DNA strand as a case study, we show that our method correctly classifies $> 94\%$ of bits. Updating the probabilities as more events are taken into consideration follows naturally from the probabilistic nature of the Bayesian approach. The focus on probabilities further allows using evidence ratios as a confidence metric for each estimate. This provides valuable information on which bits are difficult to assign and indicates the most likely sources of error. Bayesian inference can therefore inform the design of improved DNA modifications, such as patterns to facilitate identification of the correct position in the bit sequence.

While the Bayesian approach is a powerful tool for the readout of bit encodings, its generality makes it applicable to a wide range of analysis problems for nanopore data. In particular, one of its strengths lies in the computation of evidence values which probabilistically judge how well a model describes experimental data. As researchers continue to develop the physical understanding of the nanopore translocation process, competing theories can easily be assessed through Bayesian model comparison. In the context of DNA-carrier based nanopore sensing, this will shed light on open questions such as how the structure of modifications relates to the observed current drops.

\begin{acknowledgements}
	N.E. acknowledges funding from the EPSRC, Cambridge Trust and Trinity Hall, Cambridge. K.C. and U.F.K. acknowledge funding from an ERC consolidator grant (Designerpores 647144).
\end{acknowledgements}


%

\end{document}